\documentstyle[prl,aps,psfig]{revtex}
\input{epsf}
\begin{document}
\twocolumn[
\hsize\textwidth\columnwidth\hsize\csname@twocolumnfalse\endcsname
\draft
\title{``Fermi Liquid'' Shell Model Approach to Composite Fermion Excitation
Spectra in Fractional Quantum Hall States}
\author{P. Sitko$\sp{1,2}$, S. N. Yi$\sp{1,3}$, K. S. Yi$\sp{1,4}$,
 and  J. J. Quinn$\sp{1,5}$}
\address{
$\sp{1}$University of Tennessee, Knoxville, Tennessee 37969, USA
}
\address{
$\sp{2}$Institute of Physics, Technical University of Wroc{\l}aw,
Wyb. Wyspia\'nskiego 27, 50-370 Wroc{\l}aw, Poland
}
\address{
$\sp{3}$Korea Maritime University, Pusan 606-791, Korea
}
\address{
$\sp{4}$Pusan National University, Pusan 609-735, Korea
}
\address{
$\sp{5}$Oak Ridge National Laboratory, Oak Ridge, Tennessee, 37831, USA
}
\date{\today}
\maketitle

\begin{abstract}
Numerical results for the energy spectra of $N$ electrons on a spherical
surface are used as input data to determine the quasiparticle energies
and the pairwise ``Fermi liquid'' interactions of composite Fermion (CF)
excitations in fractional quantum Hall systems.
The quasiparticle energies and their interactions are then used to determine
the energy spectra, $E$ vs total angular momentum $L$, of states containing
more than two quasiparticles. The qualitative agreement  with the numerical
results gives a remarkable new confirmation of the CF picture.
\end{abstract}
]

\narrowtext

Numerical diagonalization of the interaction Hamiltonian for $N$ electrons
on a spherical surface in the presence of a radial magnetic field has
contributed greatly to our understanding of the elementary excitations in
fractional quantum Hall systems \cite{Haldane}. The energy spectra ($E$ vs
$L$, the total angular momentum) always show a lowest energy sector (or lowest
band) separated by a well-defined gap from a first excited sector \cite{He}.
In many cases the first excited sector is separated from higher states
by a slightly less well-defined gap. The object of the present paper is to
demonstrate the extent to which the low lying states can be understood in terms
of CF excitations \cite{Jain,Lopez} and their interaction energies.

We do not attempt to evaluate microscopically the quasiparticle energy (i.e.
the energy of a single CF excitation interacting with a Laughlin condensed
state) or the energy  of interaction of a quasiparticle pair. Instead we
determine these phenomenologically by using the exact numerical results
for $N$-electron systems containing zero, one or two quasiparticles.
These phenomenological ``Fermi-liquid''
 parameters are then used to study states
containing three or more quasiparticles within
a shell model analogous to that of the nucleus \cite{shell book}.
 The qualitative agreement of this
``Fermi liquid'' shell model approach
with the exact numerical spectra gives striking
new confirmation of the CF picture of the fractional quantum Hall effect.
The small quantitative discrepancies may give some clue to the relative
importance of the three body forces that arise in the Chern-Simons
transformation from electrons to composite Fermions \cite{Lopez,HLR}.

The single electron states on a Haldane sphere have eigenfunctions $|l,m>$
and eigenvalues \cite{Fano}:
\begin{equation}
\xi_l =\frac{\hbar \omega_c }{2S}(l(l+1)-S^2 )\; ,
\end{equation}
where $l$ is the angular momentum, $m$ its $z$-component, and
$2S\frac{hc}{e}$ is the strength of the magnetic monopole
which produces the radial magnetic field. The allowed values of
$l$ begin with the integral or half-integral
 value $S$ and increase in unit steps.
If we concentrate on a partially filled lowest Landau level ($l=S$),
we have $2S+1$ degenerate single particle states with $m$-values going
from $-S$ to $S$. We can form $(2S+1)![N!(2S+1-N)!]^{-1}$ antisymmetric
many body states of $N$ electrons from these degenerate single particle
states. The many body states can be written $|\alpha ,L,M>$, where $L$
and $M$ are the total angular momentum
and its $z$-component, and $\alpha$ distinguishes
different states of the same $L$.
The Wigner-Eckart theorem simplifies the diagonalization of
the interaction Hamiltonian enormously, and the final energies depend upon
$L$ and $\alpha$ but not upon $M$.

In making the transformation from electrons to composite Fermions \cite{Jain},
an even number of fictitious flux quanta are attached to each electron.
In the mean-field approximation, this leads to an effective magnetic field
$B^* $ and an effective CF filling factor $\nu^* $ which is an integer.
Since we will concentrate on states close to the Laughlin $\nu=1/3$ state
($\nu^* < 2$),
we will make use of the result that for such states $2S=3(N-1)+n_{QH}
-n_{QE}$ and $2S^* =1(N-1)+n_{QH}-n_{QE}$.
Here $n_{QH}$ ($n_{QE}$) is the number of CF quasihole (quasielectron)
excitations. $S^*$ plays exactly the same role for CF's that $S$ plays for
electrons \cite{Chen}, so that the angular momentum of a quasihole is
$l_{QH}=S^* $ and of a quasielectron (which occurs in the first excited
Landau level) $l_{QE} = S^* +1$.

The low-lying states can be classified by the number of quasiparticle
excitations they contain. The lowest energy sector contains the minimum number
consistent with the values of $N$ and $2S$. The first excited sector
contains one additional QE-QH pair.
The angular momentum of a state  is simply the total angular momentum of
the quasiparticle excitations \cite{Chen}. This can be obtained by addition
of angular momentum of the QE excitations and the QH excitations treated as
independent sets of Fermions. The energy $\xi_{QE}$ or $\xi_{QH}$ of a single
quasihole or a single quasielectron can be obtained by subtracting
the energy $E_{L=0}(2S_{\nu=1/3})$ from $E_{L=N/2}(2S=2S_{\nu=1/3}\pm 1)$.
A state containing $n_{QE}$ quasielectrons and $n_{QH}$ quasiholes
would have energy $n_{QE}\xi_{QE}+n_{QH}\xi_{QH}$ if there were no
interaction among the quasiparticle excitations. By looking at the numerical
results it is apparent that the lowest energy sector of the $n_{QE}=2$
or $n_{QH}=2$ states and the first excited sector of the Laughlin condensed
state contain a number of states  which are not degenerate. The difference
between the numerical results for the two quasiparticle state $|\alpha,L,M>$
and the sum of the bare quasiparticle energies can be interpreted as
the Fermi liquid interaction energy of the quasiparticles in that state
\cite{Yi}.
For a QE-QH pair a plot of $V_{QE-QH}(L)$ for $N=5,\;6,\;7,\;8$ electron
sytems is given in reference \cite{Yi}. For a QE-QE (or a QH-QH) pair
it is more useful to plot $V_{QE-QE}$ (or $V_{QH-QH}$) as a function
of $L_{MAX} -L$, where $L_{MAX}$ is the largest possible
angular momentum of the pair \cite{Yi,Johnson}.
The functions $V_{QP-QP'}(L)$ play the role of Landau's Fermi liquid
interaction energies \cite{Pines}. It should be noted that both
$\xi_{QP}$ and $V_{QP-QP'}(L)$ must contain effects due to both two
particle and three particle interactions. $\xi_{QP}$ is defined
as the sum of the QP kinetic energy and its interaction with the Laughlin
condensed state, which plays the role of the vacuum state.
In the CF picture this interaction contains both two body
and three body interaction terms.



Let us consider the system of $3$ particles, two of them with angular momentum
$l$, the third with angular momentum $l'$.
The wave function of the system in the state of the total angular
momentum $L$ can be written using the Racah decomposition \cite{Rose}:
\begin{equation}
\label{Racah}
|l^2 (L_1)l' L>=\sum_{L'}R_{L',L_1}|l l' (L') l L>\; .
\end{equation}
Here $|l^2 (L_1) l' L>$ denotes the state in which two QP's of angular momentum
$l$ are coupled to give a resulting two particle angular momentum $L_1 $;
$L_1 $ is then coupled with $l'$ to give a total angular momentum $L$.
In Eq. (\ref{Racah})
\[R_{L',L_1}=\sqrt{(2L_1+1)(2L'+1)} W(llLl';L_1 L')\; ,\]
where $W$ are Racah coefficients.
The matrix element of the interaction energy is
\[<l^2 (L_1)l' L|\sum_{i<k}V_{ik}|l^2 (L_2) l' L> \; .\]
We separate the interaction in the same shell ($ll$) and the interactions
between shells $ll'$ \cite{shell book}; then this matrix element becomes
\begin{equation}
\label{V2}
V_{ll}(L_1) \delta_{L_1,L_2} +
2\sum_{L'} {R^*}_{L',L_1}R_{L',L_2} V_{ll'} (L').
\end{equation}
In the case of degenerate states the interaction energies
are eigenvalues of the interaction matrix.

The problem of three or more identical particles all with  angular momentum
$l$
involves the idea of coefficients of fractional parentage.
In the special case of three particles we may write \cite{Rose}:
\[|l^3 L \alpha>=\sum_{L'}F_{L'}^{\alpha} |l^2 (L') l L>\]
where $|l^2 (L') l L>$ is the wave function of the system in which the pair
angular momentum $ll(L')$ is defined and $\alpha$
 is the additional parameter distinguishing different degenerate states.
$F_{L'}^{\alpha}$ are coefficients of fractional parentage which satisfy
the normalization condition $\sum_{L'}|F_{L'}^{\alpha}|^2 =1$.
The antisymmetry of the states leads to the
matrix equation \cite{Rose}:
$\sum_{L''}F_{L''}(A_{L'',L'}-\delta_{L'',L'})=0$ ,
where\\
$A_{L'',L'}=(-)^{3l-L+1}\sqrt{(2L''+1)(2L'+1)}W(lllL;L''L')$.

The matrix element of the interaction of the system is given by
\cite{shell book}:
\begin{equation}
\label{V1}
<l^3 L {\alpha}|\sum_{i<k}V_{ik}|l^3 L {\beta}>=
3\sum_{L'} F_{L'}^{\alpha}F_{L'}^{\beta}V_{ll}(L')\; ,
\end{equation}
where $V_{ll}(L')$ is the pair interaction of particles in $l-l$ coupling
in the state of the pair angular momentum $L'$.

The equations (\ref{V2}) and (\ref{V1})
allow us to obtain the energy of three particle
states in terms of two-particle interactions.
In our approach we make use of the quasiparticle picture of the shell
theory of the nucleus \cite{shell book}.
The main idea is that the shell with $2l+1-n$ particles can be treated in
terms of the conjugated configuration of $l^n$ quasiholes.
The allowed angular momenta of these two configurations are the same.
The idea of quasiholes
can be extended to the configuration of particles in two
shells. For example the $l^{2l}l'$ configuration can be considered
as a particle-hole pair. One can treat
the particle-hole interaction $V_{l^{2l}l'}$ as the interaction between two
particles $ll'$, $V_{ll'}$ \cite{shell book}.


Let us start by analyzing the spectrum of a three quasielectron excitation.
We keep the number of electrons fixed
and change the flux going through the sphere
($2S$); a decrease of one flux quantum from the Laughlin state
generates one quasielectron of the
angular momentum $l_{QE}=N/2$. A decrease of an
 additional flux quantum creates
two quasielectrons in the angular momentum
shell $l=N/2 -1/2$. Hence, for N=7 the allowed
angular momenta of a quasielectron pair are $5$, $3$, $1$.
In the next step we find the three-quasielectron excitations
corresponding to three CF's
in the shell $l=2.5$. The construction of antisymmetric states leads to the
states of the total angular momenta $1.5$, $2.5$, $4.5$.
In this configuration the pair angular momenta are $4$, $2$, $0$, so that
to find interaction energies of the system we have to know pairwise
interactions for the shell $l=2.5$. In fact, those can be found in the system
with a different number of electrons, in this case for $N=6$.
Another possibility is to use the data for pairwise interaction for $N=7$
assuming that quasielectron-quasielectron interaction is a function
of $RAM=L_{MAX}-L_{pair}$ \cite{Yi,Johnson}.
Here we use  the latter approach since the interactions vary with
the number of electrons. However, we believe
that the difference between using these two methods becomes unimportant
for a higher number of electrons (in the previous analysis \cite{Yi}
 very good agreement was found between pairwise interactions as a function
of $RAM$ for $N=7,8$, even though some  discrepancies occurred for $N=5,6$).

In Fig. 1A we plot numerical results of exact diagonalization (pluses) and
shell model results (circles) for three-quasielectron excitation for $N=7$.
The same is  done for $N=8$ (Fig. 1B); Fig. 2A
gives the results for
three-quasihole excitations for $N=7$, and Fig. 2B gives the same results for
$N=8$.
We observe very good qualitative agreement between numerical data
and the shell model results.

Let us now analyse the spectra of two quasielectrons (quasiholes)
and a quasihole (quasielectron).
These appear as the second excitation sector for the states containing
 one quasielectron
(quasihole). For $N=7$ the single quasielectron excitation
is in the shell $l=3.5$, so that in the second excitation sector we have
two quasielectrons in the shell $l=3.5$ and a quasihole in $l=2.5$.
The allowed angular momenta of two quasielectrons are $6,\; 4,\; 2\; ,0$,
so that the allowed total angular momenta are $0.5,\; (1.5)^2 , \;(2.5)^3 ,\;
(3.5)^3 ,\;(4.5)^3 ,\; (5.5)^2 ,\; (6.5)^2 ,\; 7.5,\; 8.5$.
Again we plot the numerical results together with the results of the shell
model (Fig. 3A). First comparison of the numerical results
with the non-interacting quasiparticles state suggests that some
of the predicted states are missing (or pushed up in energy).
In fact, we confirm
this in our shell model approach, the main reason for pushing some states
up in energy
is the large quasielectron-quasihole interaction for $L_{QE-QH}=1$.
It should be noted that other authors \cite{Dev,Rezayi,WuJain} have interpreted
the energy spectra in terms of individual CF excitations.  Although no
interactions between CF excitations were included, the overlap of the CF state
trial wavefunction with the lowest electron Landau level was evaluated.  The
absence of the $L=1$ QE-QH state in the first excited sector was attributed to
lack of overlap.
We simply assume the interaction to be large (we take the numerical
result for the lowest energy state with $L=1$ in the Laughlin state).
The same analysis is made for $N=8$ (Fig. 3B) and for two quasiholes and
a quasielectron for $N=7$ and 8 in Fig. 4A and 4B.
In all cases we observe that the qualitative feature of pushing some
states up is confirmed by the shell model approach.

There are quantitative discrepancies between the shell model results
and the numerical data. Some of them may indicate the presence of another
excited states. For example for 2QE-QH excitations we observe that an
additional
state seems to appear at $l=l_{QE} +1$, where
agreement between our model and the numerical results is the weakest.
This is undoubtly due to our neglect of
the single
quasielectron excitation in the next CF Landau level.
The ``bare'' QE energy of this excitation is very close to the ``bare''
energy of the 2QE-QH state and its angular momentum is $l_{QE} +1$.
Other discrepancies may be a measure of the role of
three particle forces which are present in the Chern-Simons approach.

It is interesting to analyse  the energy of the ground state
of $2/5$ filling. This state can be regarded as one with $n_{QE}$
quasielectrons of the $\nu =1/3$ state filling the $2l_{QE} +1$
states of the first excited CF Landau level.
The interaction energy of this closed shell configuration of $n_{QE}$
quasielectrons is given by \cite{shell book}:
\begin{equation}
V(l^{2l+1}\;L=0)=\sum_{L'}(2L'+1) V_{ll}(L').
\end{equation}
By adding this energy
to $n_{QE}\xi_{QE}$
(sum of single quasielectron energies) we can compare the resulting
energy with the exact diagonalization results.
The results for $N=6,\; 8$ are given in Table I, and they give additional
strong support for the ``Fermi liquid'' shell model picture
of CF excitations which  we use.

\acknowledgements

This work was supported in part by Oak Ridge National Laboratory, managed by
Lockheed Martin Energy Research Corp. for the US Department of Energy under
contract No. DE-AC05-96OR22464.
P.S. acknowledges  support from NATO High Technology Linkage Grant 930746
  and the Foundation for Polish Science.
S.N.Y. acknowledges support by the Korea Science and Engineering Foundation.
K.S.Y. acknowledges support by the BSRI--95--2412 program of the Ministry of
Education, Korea.
The authors thank to Drs. G. Canright and A. Kurana for helpful
discussions, particular thanks are due to Dr. X. M. Chen for his assistance
with the numerical calculations.

\begin{figure}
\vspace*{8.2cm}
\hspace*{-2.3cm}
\epsfysize=1cm
\epsfbox[0 0 30 50]{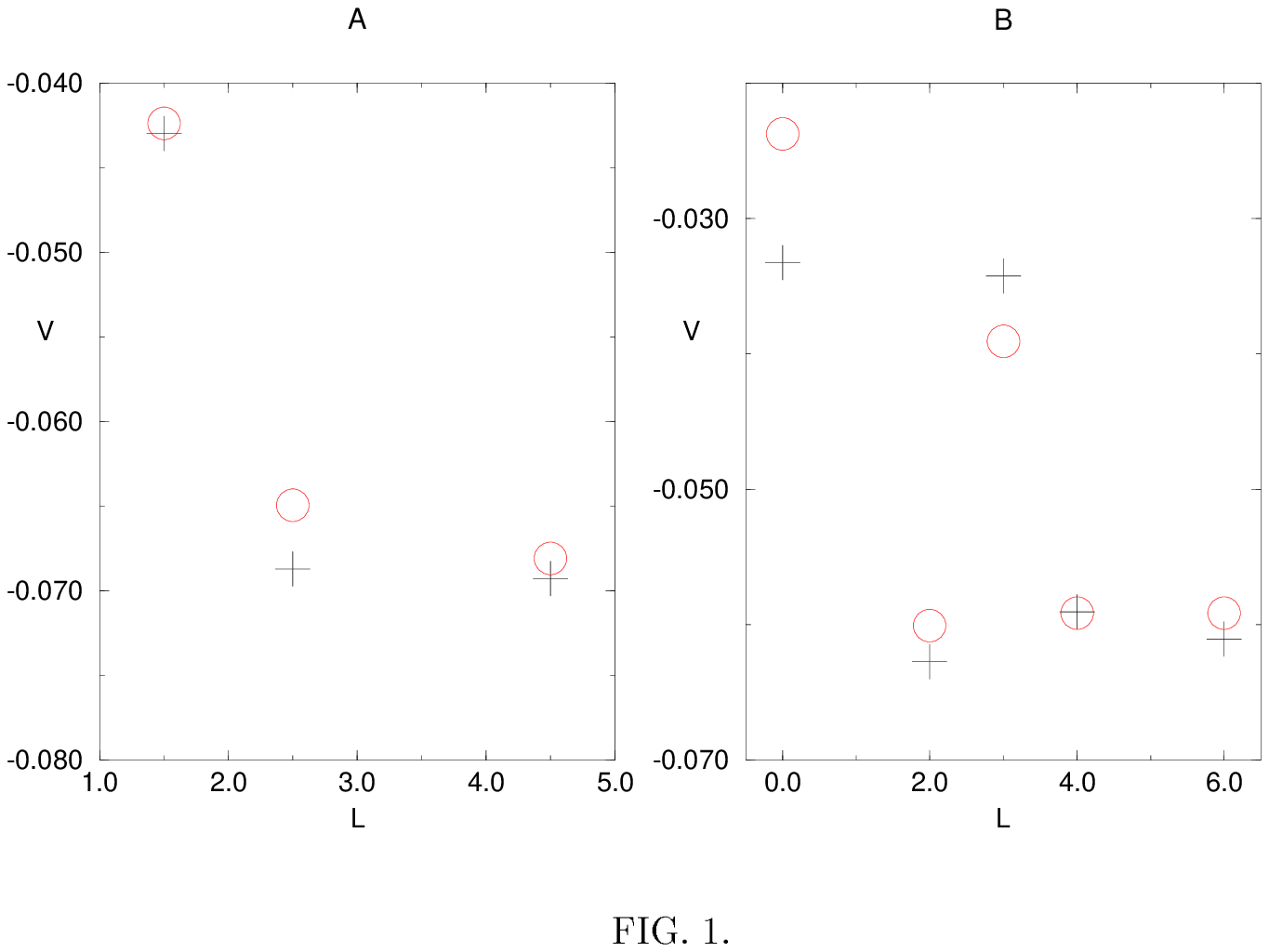}
\vspace{-4.8cm}
\caption{A. Energy spectrum of three quasielectrons for $N=7$; $l_{QE}=2.5$
and $L=1.5,\;2.5,\;4.5$. $V$ is the interaction energy, i.e. ,
$V=0$ corresponds to the ``bare'' energy  of three quasielectrons
 (in all figures
energy in units $e^2 /l_0$ where $l_0$ is the
magnetic length for the Laughlin $1/3$ state).
B. Same for $N=8$; $l_{QE}=3$ and $L=0,\;2,\;3,\;4,\;6$.}
\end{figure}

\begin{figure}
\vspace*{8.3cm}
\hspace*{-2.7cm}
\epsfysize=1cm
\epsfbox[0 0 30 50]{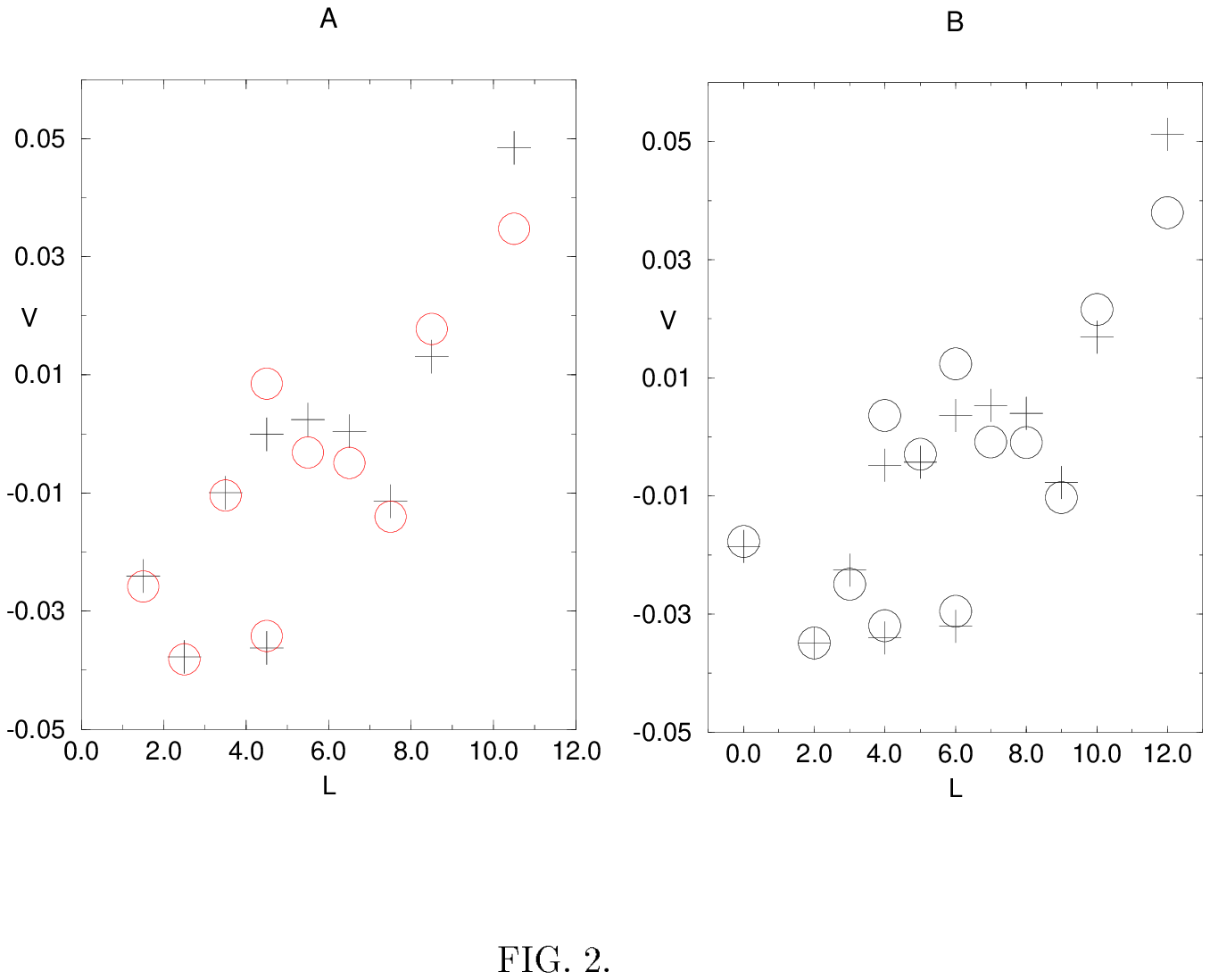}
\vspace{-4.8cm}
\caption{ A. Energy spectrum of three quasiholes for $N=7$; $l_{QH}=4.5$
and $L=1.5$, $2.5$, $3.5$, $(4.5)^2$, $5.5$,  $6.5$, $7.5$, $8.5$, $10.5$.
B. Same for $N=8$; $l_{QH}=5$ and $L=0,\;2,\;3,\;4^2,\;5$,$\;6^2,\;7,\;8,\; 9,
\;10,\;12$.}
\end{figure}

\begin{figure}
\vspace*{8.5cm}
\hspace*{-2.7cm}
\epsfysize=1cm
\epsfbox[0 0 30 50]{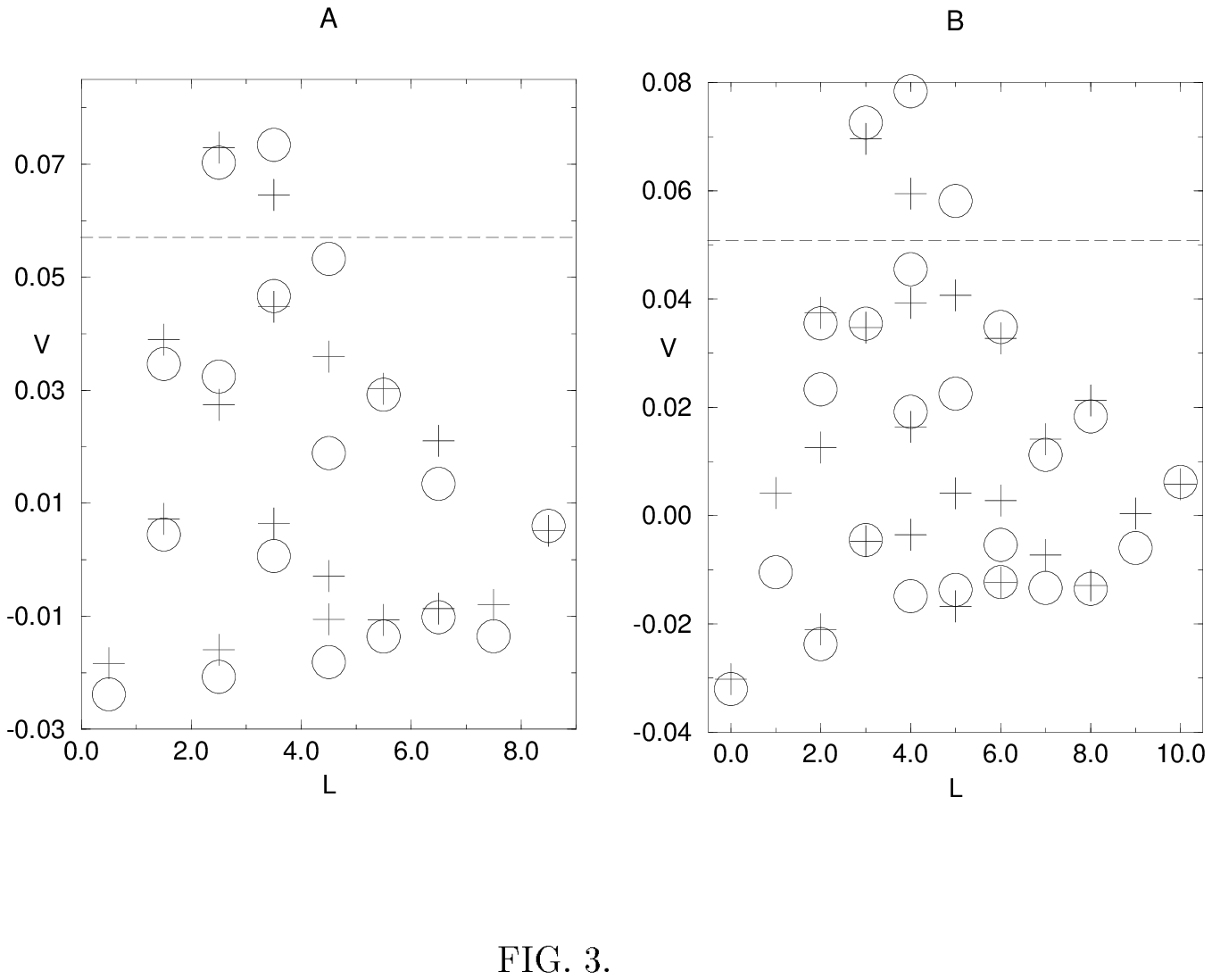}
\vspace{-4.5cm}
\caption{A. Energy spectrum of two quasielectrons and a quasihole for $N=7$;
$l_{QE}=3.5$, $l_{QH}=2.5$, and $L=0.5,\; (1.5)^2 , \;(2.5)^3 ,\;
(3.5)^3 ,\;(4.5)^3 ,\; (5.5)^2 ,\; (6.5)^2 ,\; 7.5,\; 8.5$.
The significant discrepancy at $L=4.5$ is undoubtely associated with the state
corresponding to an excited QE in the next CF Landau level which we have
neglected. The dashed line in this figure and Fig. 4 indicates the expected
position of the ``gap''.
B. Same for $N=8$;
$l_{QE}=4$, $l_{QH}=3$, and $L=0,\;1,\;2^3,\;3^3,\;4^4,\;5^3,\;6^3,\;7^2,
\;8^2,\;9,\;10$.
 The excited single QE state occurs for $L=5$.}
\end{figure}

\begin{figure}
\vspace*{8.5cm}
\hspace*{-2.7cm}
\epsfysize=1cm
\epsfbox[0 0 30 50]{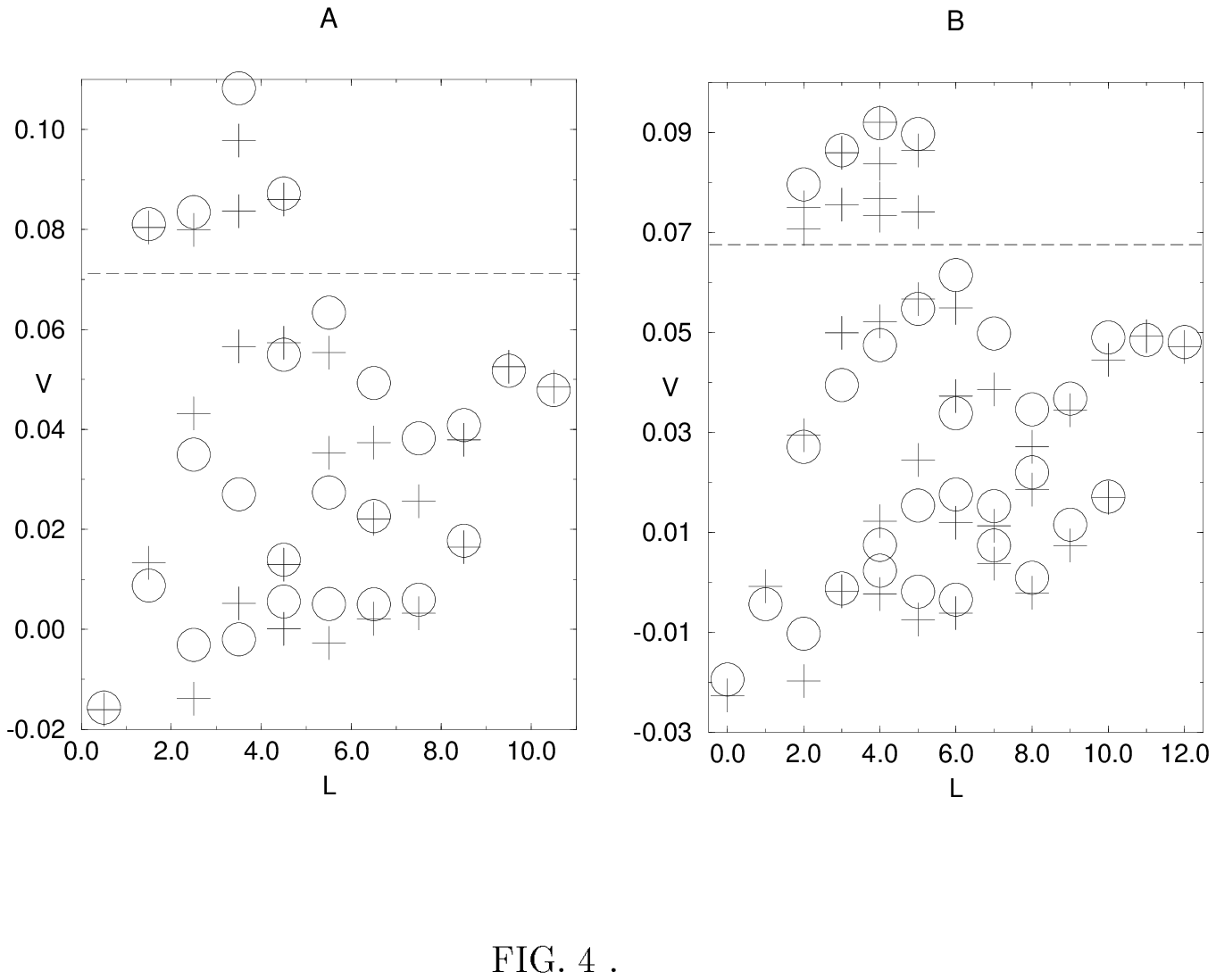}
\vspace{-4.5cm}
\caption{A. Energy spectrum of two quasiholes and a quasielectron for $N=7$;
$l_{QE}=4.5$, $l_{QH}=3.5$, and $L=0.5$, $(1.5)^2$, $(2.5)^3$, $(3.5)^3$,
$(4.5)^4$, $(5.5)^3$, $(6.5)^3$,  $(7.5)^2$, $(8.5)^2$, $9.5$,  $10.5$.
B. Same for $N=8$;
$l_{QE}=5$, $l_{QH}=4$, and $L=0,\;1,\;2^3,\;3^3,\;4^4,\;5^4,\;6^4,\;7^3,
\;8^3,\;9^2,\;10^2,\;11,\;12$.}
\end{figure}

\begin{table}
\caption{Comparison of the exact ground state energy of the state of $2/5$
with the shell model results (energy in units $e^2/l_0$ where $l_0$ is the
magnetic length for the Laughlin $1/3$ state).}
\begin{tabular}{dddd}
 &$1/3$  &\multicolumn{2}{c}{$2/5$}\\
&Exact &Exact &Predicted\\
\tableline
$N=6$ & 3.871635 & 4.001568 & 4.003656\\
$N=8$ & 6.362649 & 6.521887 & 6.535434
\end{tabular}
\end{table}

\end{document}